# Photonic Modes Prediction via Multi-Modal Diffusion Model

Jinyang Sun, [1, §] Xi Chen, [2, §] Xiumei Wang, [3] Dandan Zhu, [4†] and Xingping Zhou [5‡]

[1] *Portland Institute, Nanjing University of Posts and Telecommunications, Nanjing 210003, China*

[2] *College of Integrated Circuit Science and Engineering, Nanjing University of Posts and Telecommunications, Nanjing 210003, China*

[3] *College of Electronic and Optical Engineering, Nanjing University of Posts and Telecommunications, Nanjing 210003, China*

[4] *Institute of AI Education, Shanghai, East China Normal University, Shanghai 200333, China*

[5] *Institute of Quantum Information and Technology, and Key Lab of Broadband Wireless Communication and Sensor Network Technology, Ministry of Education, Nanjing University of Posts and Telecommunications, Nanjing 210003, China*

[†]ddzhu@mail.ecnu.edu.cn

[‡]zxp@njupt.edu.cn

*§ These authors contributed equally to this work.*

The concept of photonic modes is the cornerstone in optics and photonics, which can describe the propagation of the light. The Maxwell's equations play the role in calculating the mode field based on the structure information, while this process needs a great deal of computations, especially in the handle with a three-dimensional model. To overcome this obstacle, we introduce the Multi-Modal Diffusion model to predict the photonic modes in one certain structure. The Contrastive Language–Image Pre-training (CLIP) model is used to build the connections between photonic structures and the corresponding modes. Then we exemplify Stable Diffusion (SD) model to realize the function of optical fields generation from structure information. Our work introduces Multi-Modal deep learning to construct complex mapping between structural information and light field as high-dimensional vectors, and generates light field images based on this mapping.

    The photonic mode is one of the most common concepts in optics such as nano photonics, quantum optics and geometric optics, which is fundamental and sometimes confusing with different definitions. It can be defined as a normalized solution of Maxwell equations in vacuum [1] or the eigenfunction of an eigenproblem describing a physical system [2, 3]. The photonic mode can also be regarded as solutions for the propagation of the light or the transport of energy/information. In a specific structure, such as the cavity or waveguide, the mode is the spatial distribution of photons [4]. In fact, photonic modes can be seen as the mapping relationship between optical structure and light field distribution.

    Deep learning as a powerful tool for data analysis has been introduced into

physics, ranging from black hole detection [5], gravitational lenses [6], photonic structures design [7], quantum many-body physics, quantum computing, and chemical and material physics [8]. In optical field, the deep learning was mainly used in the design of photonic structures, such as plasmonic nanoparticles [9-11], metamaterials [12-19], photonic crystals [20-24], and integrated photonic devices [25-29]. In addition, the deep learning method can deal with the prediction of topological invariants in Hermitian and non-Hermitian systems [30-36].

Recently, Multi-Modal Machine Learning (MMML) has emerged as a novel tool in Natural Language Processing (NLP) and Computer Vision (CV) to deal with the combination of different channels of information, which is also called "multi-modal" [37, 38]. Most people associate the word modality with the sensory modalities which represent our primary channels of communication and sensation, such as vision or touch. MMML aims to build models that can process and relate information from multiple modalities. These algorithms and modeling tools show extremely good results especially for large models such as chat Generative Pre-trained Transformer (chat-GPT) [39], and visual generation models (e.g., Stable Diffusion (SD) [40]). Actually, the photonic system also behaves in the so called "multi-modal". For example, the photonic structures and the corresponding modes are two different types of modalities. Thus, we can build the connection of different modalities in photonic system via MMML, based on a database calculated by the Maxwell's equations. As we know, it requires extensive computation to get optical modes information by Maxwell's equations, and even minute parameter changes necessitate recalculations, consuming a significant amount of time and resulting in inefficiencies. However, the integration of deep learning enables models to summarize previous optical field variations, applying this knowledge to subsequent computations. This concept is akin to "transfer learning" [41] significantly reducing computational loads and enhancing efficiency by leveraging insights gained from prior calculations for future iterations.

In this work, we use MMML to generate optical fields based on the structure information. Though a dataset containing simulation optical field results and model description text, we firstly build the connections between optical structures and the corresponding optical fields by the Contrastive Language–Image Pre-training (CLIP) model [42] with fine-tuned process, which shows a great deal of improvement compared with the zero-shot network. Then we choose the SD, one of the most widely used open-source "text to image" models [40], to generate high-fidelity optical fields based on specific optical structures.

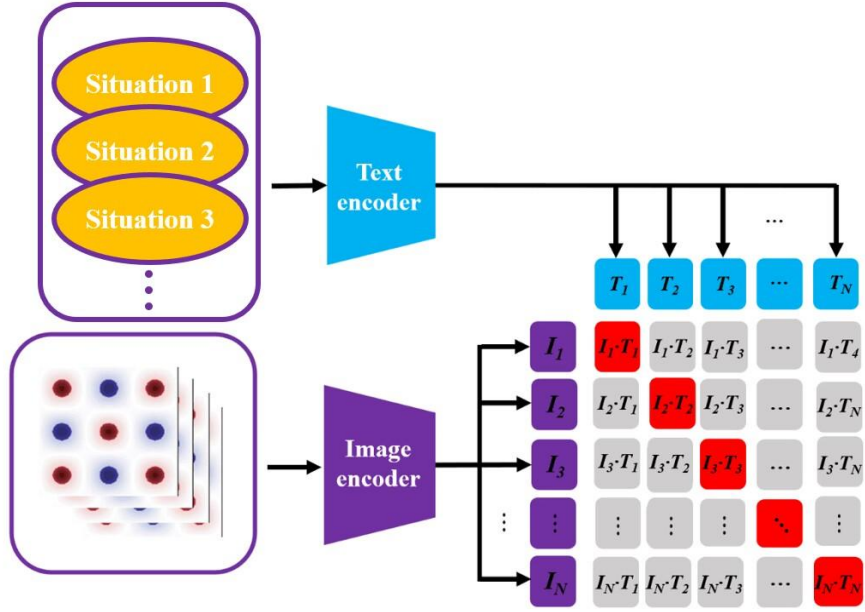

Fig.1 Schematic diagram of CLIP algorithm. The optical structure information is contained in the situation *n*.

**Results**

**CLIP modal.** To build the connections between the optical structures and their corresponding modes, we first introduce the CLIP method [42], which stands out as one of the notable landmarks in the progress of MMML. The CLIP algorithm combines an image editor and a text editor to predict the correct pairings of a batch of (image, and text) training examples as shown in Fig. 1. The image editor handles the image data and learns its feature representation, while the text editor processes the text data and learns the semantic representation of the text. Here we can regard the optical structure information as the text data and the optical field information as the image data.

The usage of our CLIP model consists of two main stages: pre-training and fine-tuning. During the pre-training stage, it trains the model using large-scale image and text datasets to establish the association between images and corresponding text. The dataset contains 400 million (image, text) pairs from publicly available sources on the Internet collected by openAI. This stage enables the model to learn the underlying relationships between different modalities. In the fine-tuning stage, we further train the model using specific task-related datasets (e.g., classification, retrieval) to enhance its performance on those specific tasks. After training, the CLIP algorithm can effectively map input images and texts, comprehend the relationship between them, and be applied to various tasks such as comparison, classification, and retrieval.

The cosine similarity is the central aspect of the CLIP algorithm. It serves as a metric for evaluating the semantic resemblance between images and text. We determine the similarity by computing the cosine similarity between the phase diagram of the light field and its corresponding parameters. The precise formula for calculating cosine similarity is presented below:

$$similarity(\boldsymbol{a},\boldsymbol{b}) = \frac{\boldsymbol{a} \cdot \boldsymbol{b}}{\|\boldsymbol{a}\| \times \|\boldsymbol{b}\|}, \tag{1}$$

where $\boldsymbol{a}$ represents the image vector extracted by the image editor, $\boldsymbol{b}$ represents the text vector extracted by the text editor, $\boldsymbol{a} \cdot \boldsymbol{b}$ represents the dot product of vectors $\boldsymbol{a}$ and $\boldsymbol{b}$, while $\|\boldsymbol{a}\|$ and $\|\boldsymbol{b}\|$ represents the modulo lengths the vector $\boldsymbol{a}$ and $\boldsymbol{b}$. Specifically, the image and text are converted into feature vectors $\boldsymbol{a}$ and $\boldsymbol{b}$ in the image encoder and text encoder, respectively, and then the cosine similarity between these vectors is calculated. If $similarity(a,b)$ is close to 1, it means that they are very close in the semantic space and have similar semantic meanings. When $similarity(a,b)$ is close to -1, it means that they are far apart in the semantic space and have opposite semantic meanings. When $similarity(a,b)$ equals to 0, it means that the two are orthogonal in the semantic space, i.e., unrelated.

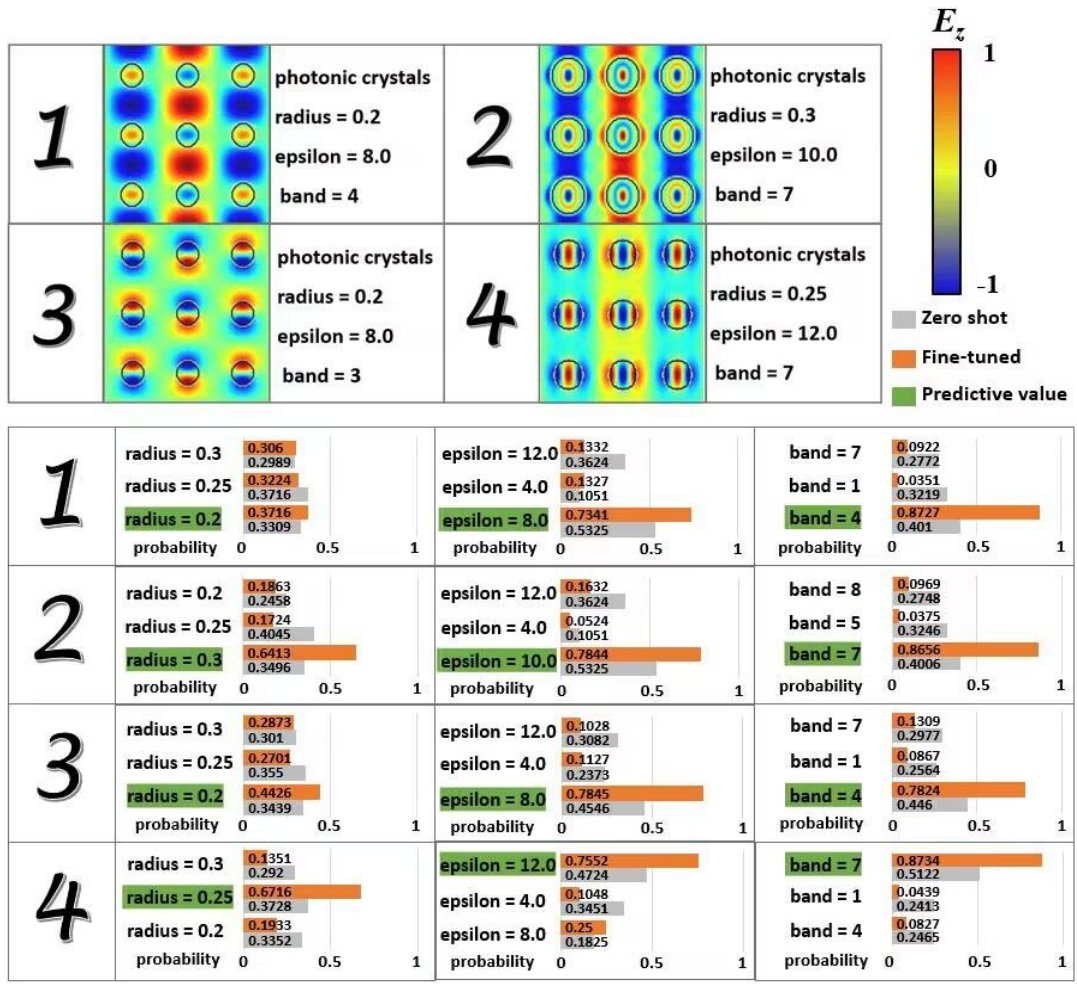

Fig. 2 Paired results of 3 × 3 photonic crystal TM ($E_z$) mode field and mode pair based on CLIP algorithm. The four images of light field distribution in the figure are numbered 1~4, and the right side of each image is part of the standard description text. For each image, we use the original CLIP network and the adjusted CLIP network to test the pairing of three different statements for three different parameters, and highlighted the pairing results given by the adjusted network in green.

**Dataset construction and testing results.** After constructing the CLIP modal, we need a comprehensive dataset comprising numerous images accompanied by corresponding text labels or descriptions in the fine-tune process. For simplicity and concreteness, we choose photonic crystals which are periodic dielectric structures exhibiting a band gap in their optical modes [43, 44], as a model to predict its optical modes. The band structure of a two-dimensional square lattice of dielectric rods in air are calculated by MPB (a free and open-source software package) [45] to construct a dataset, which includes adjustable parameters such as cylinder radius, dielectric constant, and band numbers, with Bloch wavevectors $k$ = (0.5, 0). We choose the lattice constant "*a*" of the structure, and write all distances in terms of that. In principle, any unit can be used in conjunction with MPB, provided that each parameter is expressed in a consistent unit. Therefore, we have not specified any unit in Fig. 2. As these parameters vary, the corresponding field distribution of the photonic crystal undergoes significant changes. To see more clearly, the size of our unit cell is set to 3×3. Based on the traversal parameters of the optical model, we create a dataset containing 8, 000 different simulation optical field results and model description text. Subsequently, we train the CLIP model through fine-tuning with this dataset, aiming to improve its adaptability to our optical model.

We load the optical dataset into the CLIP model and train it for 10,000 epochs, resulting in an adjusted CLIP model. To assess the effectiveness and accuracy of this trained model, we conduct tests to evaluate the correspondence between light field distribution images and optical parameter text. To this end, we randomly select four different optical field distributions with different parameters. Subsequently, we employ the trained CLIP model to examine the alignment between the images and texts, as depicted in Fig. 2. We conduct triple-text matching tests on four distinct parameterized optical field distribution images. These tests involve the radius, dielectric constant, and band number using both zero-shot and fine-tuned networks. The outcomes of these evaluations are depicted in Fig. 2, which clearly illustrates that the fine-tuned network consistently delivers more precise matching results (highlighted in green) compared to the zero-shot network.

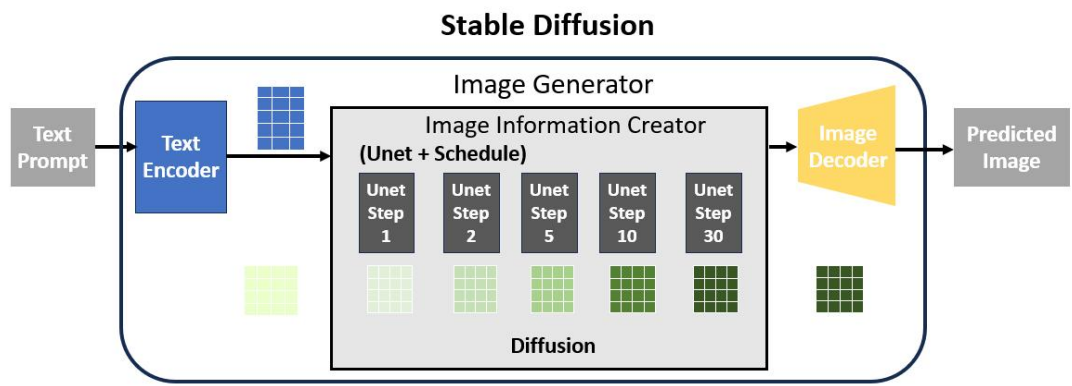

Fig. 3 Schematic diagram of SD algorithm. Enter text prompt to generate predicted image through text encoder, image generator and image decoder.

**SD modal and measurement indicators.** Following the idea of CLIP comparative learning, we predict the parameters with the greatest cosine similarity as the result of model prediction shown in the highlighted green color in Fig. 2. After validations with the theoretical values, the success rate of pairing is calculated to be 98.6%, which basically meets our requirements for the CLIP model. Then, we introduce the SD model [40] and combine it with CLIP model [42] to train the phase diagram of light field distribution. The core idea of SD is that each image has a certain regular distribution Thus, the distribution information contained in the text can be used to guide the denoising process and generate an image matching the text information. In order to make the input text information become the machine information that SD model can understand, we need to give SD model a "bridge" between text information and machine data information namely CLIP text encoder model, as shown in Fig. 3.

For the SD algorithm, it is an image denoising method based on partial differential equation. First, we add Gaussian noise to the image: $I_n(x,y) = I(x,y) + G(\mu, \sigma^2)$, where $I(x,y)$ represents the pixel coordinates in the original image, $G(\mu, \sigma^2)$ is average $\mu$ and variance $\sigma^2$ Gaussian random variable, $I_n(x,y)$ is the image after adding noise. Then, we perform stable diffusion denoising on the noisy image by solving the nonlinear diffusion equation [40]:

$$\frac{\partial I(x,y,t)}{\partial t} = \nabla \cdot (c(\|\nabla I(x,y,t)\|)\nabla I(x,y,t)), \qquad (2)$$

where $I(x,y,t)$ represents the image position coordinates $(x,y)$ and the brightness value $t$, $\nabla$ is the gradient operator, and $c(\|\nabla I(x,y,t)\|)$ is the diffusion coefficient, which is a function of gradient intensity. By iteratively solving the above partial differential equation, the noise in the image can be gradually reduced and the edge information of the image can be retained. In order to balance the trade-off between denoising and edge preserving, it is necessary to select an appropriate diffusion coefficient $c(\|\nabla I(x,y,t)\|)$. Here we choose a common diffusion coefficient based on Perona-Malik model [40]:

$$c(\|\nabla I(x,y,t)\|) = e^{-(\frac{\|\nabla I(x,y,t)\|}{K})^2}, \qquad (3)$$

where $K$ is the parameter that controls the de-noising degree, $\|\cdots\|$ represents square of Euclidean distance. In our work, we add Gaussian noise to the original image and use the stable diffusion method by solving Eq. (2) to remove the noise, so as to obtain the denoised image. We have also provided an explanation of its working principle in APPENDIX A.

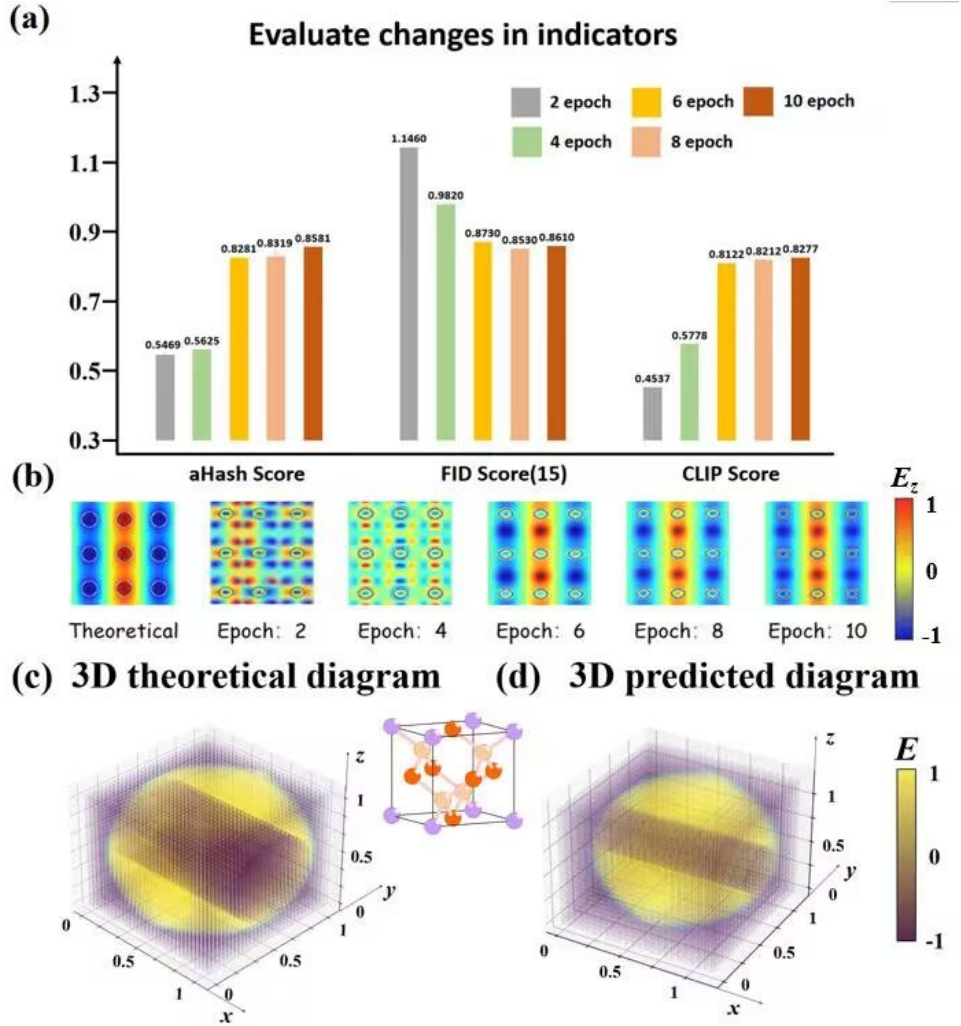

Fig. 4 (a) Histogram of three indicators at different epoch stages. (b) Comparison chart of a group of theoretical prediction images at different epoch stages. (c-d) Theoretical and predicted diagram of a 3D spherical lattice light field distribution.

In order to measure the effect of network generated images accurately, we introduced three measurement indicators: Average Hashing (aHash) score [46], Fréchet Inception Distance (FID) score [47] and CLIP score [42]. The aHash algorithm is an image similarity comparison method based on Perceptual Hashing technology. We take the aHash similarity to measure the similarity between the light field distribution map generated by SD model and the theoretical light field distribution map. The calculation process is as follows:

First, the two input images are converted into gray-scale images respectively, and their average values can be calculated as

$$G(I) = \frac{1}{W \times H} \sum_{i=1}^{H} \sum_{j=1}^{W} I(i,j), \tag{4}$$

where *I(i,j)* refers to graying out the pixel values in row *i* and column *j* of the image, *W* is the width of the image, and *H* is the height of the image. Next, each pixel in the image is traversed, and the binary value is used to represent the pixel. The specific expression is as follows:

$$B(I, G(I)) = \begin{cases} 1 & I(i,j) \geq G(I) \\ 0 & I(i,j) < G(I) \end{cases} \quad i \in [0,H), j \in [0,W) \tag{5}$$

where *I(i,j)* represents the gray value of the image in row *i* and column *j*, *G(I)* represents the average gray value of the image, and *B(I,G(I))* is the binary feature vector of the image. Finally, the binary vectors of the two images are compared, and the number of the same elements is compared with the vector length to obtain the similarity between the original image and the generated image. Therefore, aHash score is a relatively simple and intuitive judgment index.

FID score is a metric used to evaluate the quality of generated images. It is specifically used to evaluate the performance of images generated by the model. Its calculation formula is as follows:

$$FID(\boldsymbol{P},\boldsymbol{Q}) = \|\boldsymbol{u}_P - \boldsymbol{u}_Q\|^2 + Tr(\boldsymbol{C}_P + \boldsymbol{C}_Q - 2\sqrt{\boldsymbol{C}_P \boldsymbol{C}_Q}), \tag{6}$$

where $\boldsymbol{P}$ represents the distribution of theoretical image, $\boldsymbol{Q}$ represents the distribution of the generated image, $\boldsymbol{u}_P$ and $\boldsymbol{u}_Q$ represents the eigenvectors of images P and Q respectively, $\boldsymbol{C}_p$ and $\boldsymbol{C}_Q$ represents the covariance matrix of the eigenvectors of two distributions respectively, $Tr()$ denotes the trace operation of the matrix, and $\|\cdots\|^2$ denotes the square of the Euclide. FID score can evaluate the difference between the generated model and the real data distribution. The lower the value, the closer the generated image is to the real image. Due to the significantly larger values of the FID Score compared to the aHash Score and CLIP Score, we extract a coefficient of 15 from the FID Score to adjust the numerical scale, as depicted in Fig. 4(a).

CLIP score is a measure of the relevance between text and image. It measures the similarity between input text and image by converting them into feature vectors through CLIP model, and then calculating the cosine similarity between them. The calculation formula is as follows:

$$CLIP - S(\boldsymbol{c},\boldsymbol{v}) = w \times \max(\cos(\boldsymbol{c},\boldsymbol{v}), 0), \tag{7}$$

where $\boldsymbol{c}$ represents the input text data, $\boldsymbol{v}$ represents the input predictive image data, and $w$ is the coefficient parameter. Here we set $w = 1$. The higher the CLIP score, the higher the correlation between image text pairs. Therefore, when the CLIP score of this model is closer to 1, the image text correlation is higher and higher.

Here we list a group of fields predicted during different epochs, and take use of the above three indicators to evaluate the effect of networks with different training epochs, as shown in Fig. 4(a). The parameters of this model are set as radius = 0.2, epsilon = 9.5, Bloch wavevectors $k = (0.5, 0)$, and band = 7 with the same type photonic crystals in Fig. 2. We write all distances in terms of the lattice constant $a$. The closer the aHash similarity is to 1, the more the generated image resembles the theoretical light field distribution map. Similarly, a smaller FID score indicates that the generated image is closer to the theoretical light field distribution map. A CLIP score closer to 1 indicates that the generated light field distribution map better matches the textual description.

Based on the three measurement indicators, it is evident that the aHash value, FID score, and CLIP score of the network display signs of stabilization, as the model undergoes training until the 6$^{th}$ epoch. It indicates that more training of the network yields minimal improvement, suggesting that the model tends to reach a stable state. Additionally, it demonstrates a significant enhancement in the quality of generated images from epochs 2 to 6 in Fig. 4(b). However, beyond the 6$^{th}$ epoch, no noticeable differences can be observed in the images generated by our model. In this case, we offer an example of the training process, and we've included some other light field prediction images in APPENDIX B and the prediction process is provided in APPENDIX C.

Additionally, we also calculate the three-dimensional optical field distribution of the lattice by MPB shown in Fig. 4(c-d). Our model not only has the ability to predict two-dimensional field distributions but also has the ability of handling three-dimensional light field prediction tasks with specific training. For this investigation, we employ the following parameter values: radius = 0.4, epsilon = 9.2, and band = 5 with Bloch wavevectors $k = (0, 0.625, 0.375)$, and subsequently predict the three-dimensional light field. By comparing our prediction results with the theoretical light field distribution, we observe that the fundamental features are still present in the predicted image, of which the aHash score, the FID score and the FID score are 0.8224, 13.7415, and 0.7728, respectively. As our dataset obtained by MPB is accurate and complete, a larger dataset can improve the accuracy of prediction. Furthermore, we provide a way to improve the details of photonic modes shown in APPENDIX D.

**Discussion**

To summarize, we have introduced the MMML to photonic modes prediction. We show that the photonic crystal mode can be well matched and predicted by the CLIP and the SD algorithm based on a dataset consisted by paired optical structure and field information. Our method can significantly reduce the computational complexity of optical structures, especially in the mission of light field optimization. We have successfully generated optical fields based on the structure information utilizing the "text to image" algorithm. Actually, as the optical structure can be set as the input image data, the "image to image" generative models [49] can also be applied to find the relation between optical structure and the corresponding modes. Moreover, the "any to any" generative models [50] can help to generate target modal from input combinations of any set of input conditions, as the design of photonic structures needs more than one the so-called modality or requirement, such as basic optical structures, target optical fields, and optical spectrum. We believe that the MMML also can be extended to other physical platforms, such as mechanism, acoustics, electrical circuit and so on, duo to the generalization of the "multi-modal" concept.

**Acknowledgements**

The authors thank for the support by National Natural Science Foundation of China under (Grant 62001289), NUPTSF (Grants No. NY220119, NY221055). We thank Professor Xiaofei Li for useful discussions.

# APPENDIX A: COMBINED FLOWCHART OF SD AND CLIP

*SD and CLIP combined search map.* SD combined with CLIP forms an efficient method for text and image matching. By employing text encoders and image encoders, texts and images from different contexts are transformed into high-dimensional vector representations. The similarity between these vectors in the semantic space is measured using dot product calculations, which effectively captures the correlation between texts and images

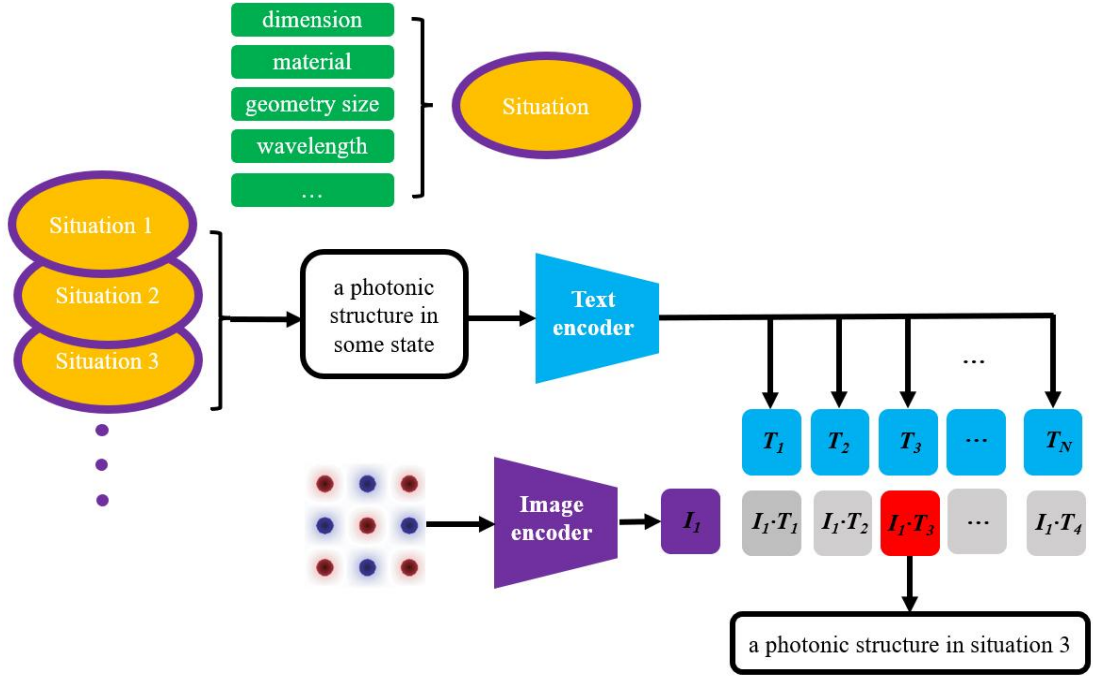

Fig. S1 SD and CLIP combined search map. Different situation texts and images are transformed into high-dimensional vectors by text encoder and image encoder to calculate the dot product, and then find the best image text matching combination according to the calculation results.

The aim of the search mapping process is to identify the best matching combination of images and texts that possess the highest similarity scores in the semantic space. By comparing the similarity scores among different pairs of images and texts, the optimal match can be determined. This matching procedure facilitates accurate image description or image classification tasks shown in Fig. S1.

In the fine-tuning process, we select the pre-training model vit-b/32 released by OpenAI as the foundation. The image processor of this model takes traditional convolutional filters with self-attention mechanism calculated with the 7×7 image block grids. The text processor is similarly represented by a 12-layer transformer trained over a vocab of 49K BPE token types. Both the text and image networks output a 512-D vector, respectively. The specific pipeline of the CLIP network is shown in Fig. 1, where the text encoder is used to extract the feature vector $I$ of the text information, and the image encoder is used to extract the feature vector $T$ of the image. The contrast loss of the CLIP algorithm enables the model to establish a connection between the input image and text by minimizing the distance between similar images and text pairs, and maximizing the distance between dissimilar pairs.

# APPENDIX B: ADDITIONAL PREDICTED IMAGES

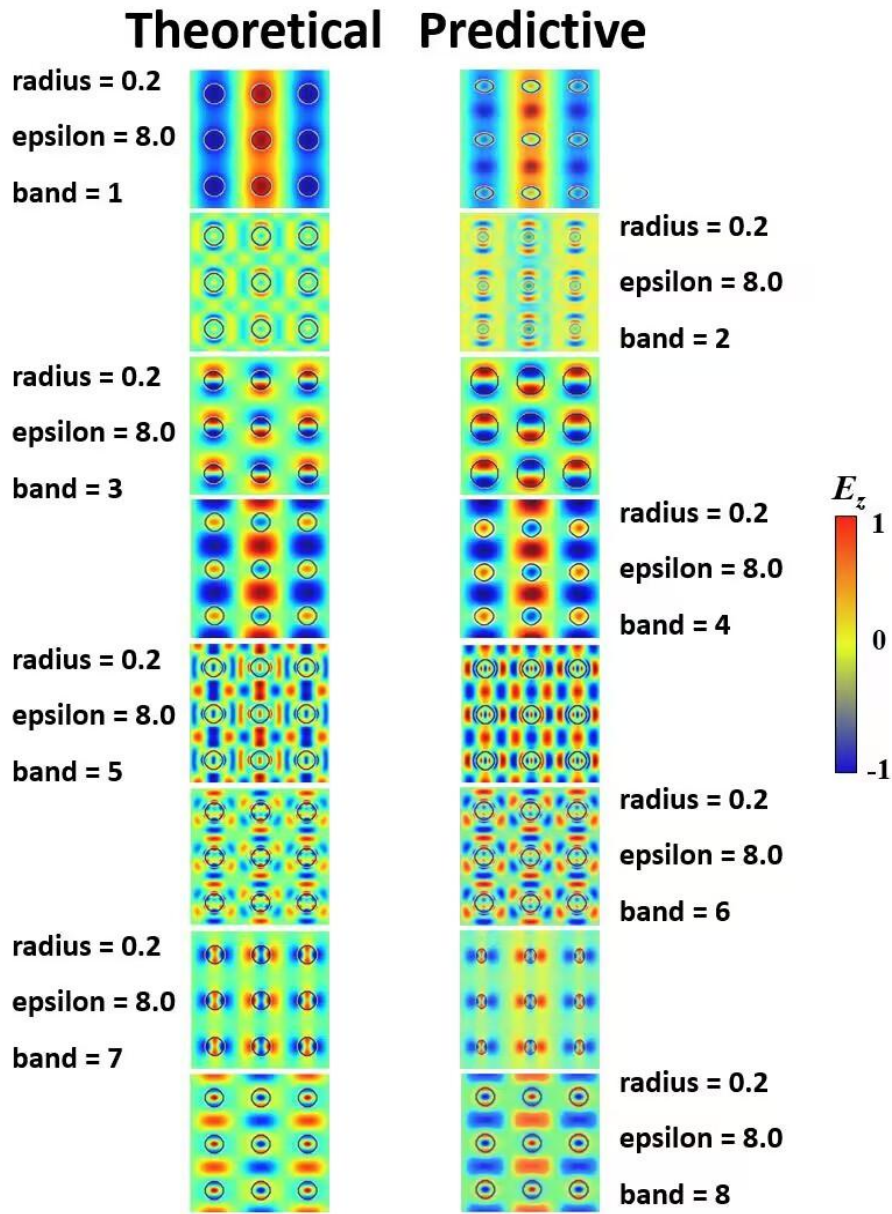

Fig. S2 Comparison chart of band scan theory and prediction. The figure shows the comparison between the theory and prediction of fixed radius and epsilon parameters with bands varying from 1 to 8.

We have provided additional theoretical and predicted images of the light field distribution for different bands in Fig. S2. We present a series of predicted light field images for bands 1 to 8 under radius = 0.2, epsilon = 8.0 as an example to demonstrate the performance of the network's predictions.

# APPENDIX C: PREDICTION RESULTS OF A SET OF SD MODELS UNDER DIFFERENT ADDINET WEIGHTS AND EPOCHS

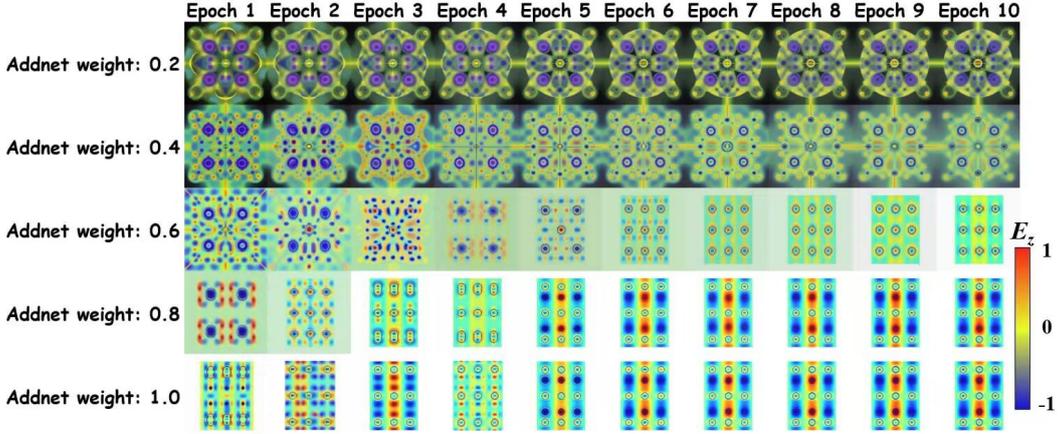

Fig. S3 Prediction results of a trained SD model under different Addnet weights and epochs.

We provide the detailed prediction process in Fig. 4(b), demonstrating the predictions of our trained SD model on the theoretical image under different Addnet weights and epochs. Addnet weight represents the degree of intervention by the trained network model in predicting images. It is evident that as the Addnet weight increases, the image prediction quality improves. In fact, the final prediction corresponds to Addnet weight = 1.0, Epoch 10. We present the images of the model training process here to intuitively understand the predictive performance and convergence of the model. Convincingly, the results indicate that the model's performance aligns with the theoretical expectations.

The text encoder in the SD model encodes input text information, generating a characteristic matrix for understanding. This matrix is then used by the image generator, consisting of a U-Net network and Schedule algorithm [40]. The U-Net network predicts noise residuals, reconstructs the matrix, converting noise into the latent feature. The Schedule algorithm optimizes noise, controlling intensity and coordinating the generation process. In SD, around 50 to 100 iterative steps refine the latent feature, denoise and increasing semantic information. Finally, the optimized latent Feature will be put into the image decoder, reconstructing it into a pixel image.

To enhance convergence and performance during training, we utilize 10 epochs with 20 steps each. Besides, we employ Adam as the optimizer, cosine with restarts as the LR scheduler [40], and implement a 10% LR warm-up to improve stability and generalization. Moreover, to alleviate video memory constraints, we incorporate gradient checkpointing and leverage shuffle caption and XFormer technologies as well.

In the training process, we take 512 × 512 pixel images as the input, and preprocess the dataset in detail, including standardization, image clipping, and label processing. In order to further optimize the training, we select the appropriate learning rate, in which the learning rate of text encoder is 0.00005, and the learning rate of U-Net is 0.0001. Our results show that after several epoch training, the predicted phase diagram of light field distribution has achieved satisfactory results.

# APPENDIX D: PREDICTION RESULTS OF A SET OF SD MODELS UNDER DIFFERENT ADDINET WEIGHTS AND EPOCHS

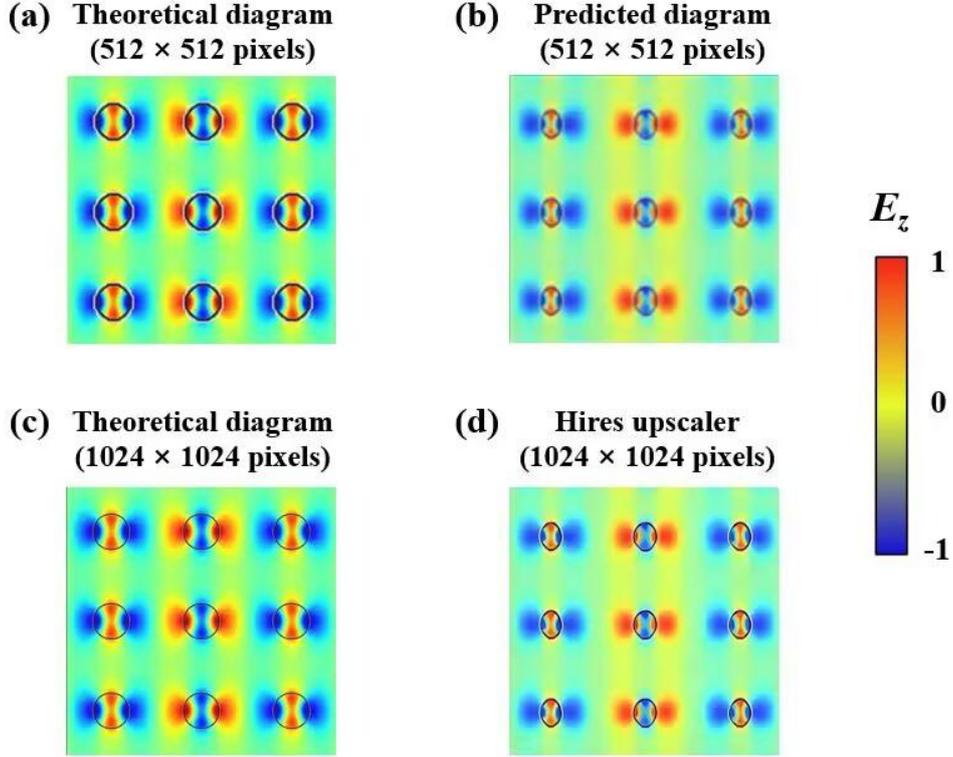

Fig. S4 The super-resolution function of SD algorithm for predicting spatial light field results. (a)-(b) Theory and prediction diagrams for a resolution of 512× 512 pixels. (c)-(d) Theory and prediction diagrams for a resolution of 1024× 1024 pixels.

    We provide a way towards enhancing the super-resolution capability of the SD algorithm in predicting light field results at higher resolutions. Since high-precision light field image calculation requires a lot of computing resources, we choose to use some algorithms to reduce the amount of calculation. Latent algorithm is the image enhancement technique based on the Variational Autoencoder (VAE) model [48]. By encoding the original image into latent vectors and subsequently reconstructing them through random sampling, we can effectively improve the quality, contrast, and clarity of the image.

    To accomplish this, we employ a reverse diffusion process in conjunction with Eq. 2, enabling us to generate images with higher pixel values. Thus, we select a specific group of images for demonstration purposes, as presented in Fig. S4. The parameters are set as radius = 0.2, epsilon = 9.5, and band = 7 with Bloch wavevectors $k$ = (0.5, 0) in terms of the lattice constant $a$. In this process, the original input image is shown in

Fig. S4(a), and we represent the predicted image generated by our trained SD model in Fig. S4(b). Next, we employ the latent algorithm [40] to reconstruct the image and generate a high-resolution image depicted in Fig. S4(d), whose theoretical image is shown in Fig. 5(c). Then, we take the aHash score to evaluate the effectiveness of these improvements. The results indicate that we achieve an aHash score of 0.86 in Fig. S4 (b), while an aHash score of 0.92 is got in Fig. S4(d). It is evident that our algorithm notably can improve image precision and accuracy, holding considerable significance in computational high-precision light field imaging.